\documentclass[10.5pt, conference]{IEEEtran}

\IEEEoverridecommandlockouts
\usepackage{cite}
\usepackage{amsmath,amssymb,amsfonts}
\usepackage[dvips]{graphicx}

\usepackage{algorithmic}
\usepackage{graphicx}
\usepackage{textcomp}
\usepackage{xcolor}
\usepackage[ruled,vlined,linesnumbered,noresetcount]{algorithm2e}

\def\BibTeX{{\rm B\kern-.05em{\sc i\kern-.025em b}\kern-.08em
    T\kern-.1667em\lower.7ex\hbox{E}\kern-.125emX}}
\begin{document}

\title{A Framework for Prediction and Storage of Battery Life in IoT Devices using DNN and Blockchain\\
}

\author{\IEEEauthorblockN{1\textsuperscript{st} Siva Rama Krishnan Somayaji
}
\IEEEauthorblockA{\textit{School of Information Technology} \\
\textit{Vellore Institute of Technology}\\
Vellore, India \\
siva.s@vit.ac.in}
\and
\IEEEauthorblockN{2\textsuperscript{nd} Mamoun Alazab}
\IEEEauthorblockA{\textit{College of Engineering, IT and Environment} \\
\textit{Charles Darwin University}\\
0909, Australia \\
mamoun.alazab@cdu.edu.au}
\and
\IEEEauthorblockN{3\textsuperscript{rd} Manoj MK}
\IEEEauthorblockA{\textit{School of Information Technology} \\
	\textit{Vellore Institute of Technology}\\
	Vellore, India \\
	mk.manoj2015@vit.ac.in}
\and
\IEEEauthorblockN{4\textsuperscript{th} Antonio Bucchiarone}
\IEEEauthorblockA{\textit{Fondazione Bruno Kessler (FBK)} \\
	\textit{Trento}\\
	Italy\\
	bucchiarone@fbk.eu}
\and
\IEEEauthorblockN{5\textsuperscript{th} Chiranji Lal Chowdhary}
\IEEEauthorblockA{\textit{School of Information Technology} \\
	\textit{Vellore Institute of Technology}\\
	Vellore, India \\
	chiranji.lal@vit.ac.in}
\and
\IEEEauthorblockN{6\textsuperscript{th} Thippa Reddy Gadekallu}
\IEEEauthorblockA{\textit{School of Information Technology} \\
	\textit{Vellore Institute of Technology}\\
	Vellore, India \\
	thippareddy.g@vit.ac.in}

}

\maketitle

\begin{abstract}
As digitization increases, the need to automate various entities becomes crucial for development. The data generated by the IoT devices need to be processed accurately and in a secure manner. The basis for the success of such a scenario requires blockchain as a means of unalterable data storage to improve the overall security and trust in the system. By providing trust in an automated system, with real-time data updates to all stakeholders, an improved form of implementation takes the stage and can help reduce the stress of adaptability to complete automated systems. This research focuses on a use case with respect to the real time Internet of Things (IoT) network which is deployed at the beach of Chicago Park District. This real time data which is collected from various sensors is then used to design a predictive model using Deep Neural Networks for estimating the battery life of IoT sensors that is deployed at the beach. This proposed model could help the government to plan for placing orders of replaceable batteries before time so that there can be an uninterrupted service. Since this data is sensitive and requires to be secured, the predicted battery life value is stored in blockchain which would be a tamper-proof record of the data.
\end{abstract}


\section{Introduction}
\label{sec:introduction}
The recent era is moving towards technology-based solutions for real-world problems. The intervention of technology in day to day activities is critically high. One of the trending Information and Communication technology is Internet of Things (IoT) which is the widely used in the recent decade. IOT is a term that connects people worldwide with places, products, things, etc, \cite{iwendi2020metaheuristic}. 
This also provides a wide opportunity for value creation and capture. The IoT networks which are designed are deployed in the real-world environment and critical data are transmitted to the cloud \cite{ray2016survey} where the data is stored. Though the world is towards this powerful and enabling technology, the deployment of such IoT networks is limited and it is not practiced in large numbers in the real environment. The reason for this hesitation in deployment is twofold. The first being battery usage of the sensor and the rate at which the battery drains out. Once the battery drains out, it needs to be replaced. During this time span, critical data which may affect the overall analysis and the decisions making process may be lost and this result in the reduction of profits for the organization. The second major hinderance is security of the sensitive data. Since the data is transmitted via the internet to the cloud, there is a chance for an attack in between the transmission and also in the storage area. There is a chance of huge threat to the IoT Network as well as the Recommender Systems \cite{khan2018iot}.

As per a recent forecast by Gartner’s, the IoT networks \cite{rahman2016securing}  and their deployment in the real world are expected to have a growth of around 32\% by 2021 which results to be around 25.1 billion IoT units installed. This growth must be accompanied by continued research on the security and scalability issues of Iot networks. The blockchain or Distributed Ledger technology can be a viable solution to the scalability and security concerns of Iot devices \cite{khan2018iot, chen2019collaborative, chaudhry2018enabling,tariq2019security, khalid2020decentralized, hakak2020securing}. Blockchain works on the concept of shared digital ledger that is continually recompiled to all the users of the network for every transaction made. Once the transactions are carried out, these transactions are validated and then recorded into the digital ledger which cannot be modified or deleted. The right to view, amend or transact newer information into the ledger is a benefit given to certain individuals of the network in the community.

Motivation for this work:

\begin{itemize}
	\item The battery life depletion is a major problem in real time deployed sensor networks. It can be the major contributing factor to why analytic predictions made with the data could not be met
	\item Mission critical data could be lost during the timespan of the battery replacement.The dependence of batteries are a major inhibiting factor for the depletion of growth rate in IoT sensor deployment.
	\item It can also be seen that the labor cost for monitoring and replacing the batteries overtake the benefits provided by the deployed sensors themselves.
	\item By having a third party take control over the actual task of battery replacement, the trust on the entity could be the only assurance for their actions. A decentralized ledger system could provide an immutable copy of the predicted battery life thereby combating the possibilities of human error or scam
	
\end{itemize}

The main contributions of this work are:

\begin{itemize}
	\item A prediction model is proposed using Deep Neural Networks for predicting the battery life of the sensors deployed at six locations in the beach water of Chicago Park District.
		\item Securing the sensitive predicted battery life values by storing them in the blockchain to attain a tamper proof record of this data.
\end{itemize}

The rest of the paper is organized as follows. Section \ref{sec2} discusses the existing literature done on application related to Intelligent IoT with blockchain. Section \ref{sec3} describes the background of blockchain technology and proposed architecture. Section \ref{sec5} discusses the results and comparative analysis. Section \ref{sec6} highlights the conclusion and future work.
\section{LITERATURE REVIEW}
\label{sec2}

K. Kostal and K. Salah \cite{kovstal2019management} used a private blockchain for IoT device configuration management system in CISCO networks platforms. It is tested in two ways: firstly, the utility of the proposed system design is verified by a functional scenario.  In the second phase, it is used to estimate download time, accept and install time of the updated configuration file. This system was run on CISCO TCL scripts. J. Lee \cite{lee2018patch} used a blockchain patch-transporter, an approach to update software in IoT Platforms in a peer-peer manner. In this method, a blockchain based smart contract is used for validating transportation patches and enticement. P. Helebrandt et. al. \cite{helebrandt2018blockchain} introduced private blockchain-based architecture to provide security in enterprise-level in the creation of the trust system among distributed communication parties to resolve security issues. This private blockchain system is developed to defend unauthorized data creation and improvise data integrity.

C. Qu et. al. \cite{qu2018blockchain} proposed an IoT credibility verification method based on self-organized blockchain structures of layers and communication parties in order to provide security for various data-intensive applications running on different machines. This method has the advantages of storage space and response time. However, the challenge remains for the scenario of large scale IoT platforms to determine how to choose and control the number of blockchain structures (BCS) and the height of the tree, respectively.  U. Guin et. al. \cite{guin2018ensuring} developed a blockchain architecture based on physically unclonable functions (PUFs) to counter the counterfeit, addressing the clone problems and reliability of the supply chain. This would also provide edge device authentication. The PUFs are used to generate a unique identification number for edge devices registered at manufacturers through hash functions. Registered edge devices data will be stored and will be globally available in blockchain architecture to verify the edge device by anyone anywhere without knowledge of manufactures.  

O. Novo et. al. \cite{novo2018scalable} proposed a proof-of-concept based on blockchain decentralized system for large distributed IoT systems to provide scalable access management. Smart contact contains set operations such as revoke permission, Query Manager, Query Permission, remove and add manager. N. Fotiou et. al. \cite{fotiou2019secure} designed an Ethereum based blockchain architecture to control a group of IoT devices. This system uses custom blockchain tokens to build efficient and robust access control methods. X. Liang et. al. \cite{liang2017towards} proposed public blockchain architecture along with cloud server to secure drone communication, the transmission of data, preserving data integrity and cloud auditing. The drone chain contains the registration of drone, data transmission, blockchain receipt, data validation in cloud and decision making.  This system generates a hash code for data records collected from multiple drones. The system reduces the drone’s communication burden. 

A. Outchakoucht et. al. \cite{outchakoucht2017dynamic} proposed a blockchain framework to provide an effective, scalable, and self-adjusted security policy with the distributed aspect which is strongly recommended in the IoT and machine learning algorithms. It ends with an explicit inference system designed for better understanding. E. Karafiloski, and A. Mishev \cite{karafiloski2017blockchain} emphasizes the importance of Blockchain technology to some of the Big Data areas in providing solutions. The key aspects proposed by the authors are related to user authentication, user-specific restriction access, data protection and user data record keeping. 

J. Ali et. al. \cite{ali2020towards} developed a custom Behaviour Monitor which extracts the activity of each device for analyzing their behaviour with the use of machine learning algorithms. The authors also proposed Trusted Execution Technology (TEE) to formulate a secure execution environment for sensitive application code and data on the blockchain.

The papers studied in this section along with the key findings and their limitations is summarized in Table \ref{Table0}.

\begin{table*}[h!]
	\centering
	\caption{Summary of significant works}
	\label{Table0}
	\begin{tabular}{|p{15pt}|p{110pt}|p{130pt}|p{125pt}|}
		\hline
		Ref. & Methods & Observations & Research Gaps  \\
		\hline
		
		\cite{kovstal2019management} & Chain code-CRUD:	
		Create, 
		Read, 
		Update, 
		Delete. 
		&
		Improved management Monitoring of IoT devices using a private BC &
		
		Low Accuracy, 
		Moderately secure, 
		\\ \hline
		
		\cite{lee2018patch} &
		Patch Transporter:  
		Faithful Delivery,  
		Authenticated Origin. &
		For delivering patches through an incentive system &
		No specification for architecture,  
		Accuracy.\\  \hline
		
		\cite{helebrandt2018blockchain} &
		Introduced architecture for monitoring and managing enterprise  networks through private blockchain &
		Distributed BC removes single point of failure  &
		Accuracy,  
		Tampering, 
		\\ \hline
		\cite{guin2018ensuring} &
		SRAM-based physically, unclonable functions (PUFs), Light Weight, Mining Algorithm &
		Counterfeit, clone problems, reliability of supply chain, provides edge device authentication. &
		Can be used as local block chain only \par
		Cannot be scaled \\  \hline
		
		\cite{novo2018scalable} &
		Smart contact  set operations:\par
		Revoke permission 
		Query manager 
		Query Permission 
		Remove manager
		Add manager 
		&
		A proof-of-concept architecture for IoT &
		Accuracy, 
		Tampering.   \\ \hline
		\cite{fotiou2019secure} &
		Blockchains, smart contracts &
		Distributed nature build an event-based system for managing IoT devices connected to Web of Things gateways &
		Less Accuracy,
		Efficiency. 
		Tampering.   \\		\hline
	\end{tabular}
\end{table*}

\section{Proposed Blockchain based Prediction Model}
\label{sec3}
\begin{figure}[h]
	\centering
	\includegraphics[width=8cm]{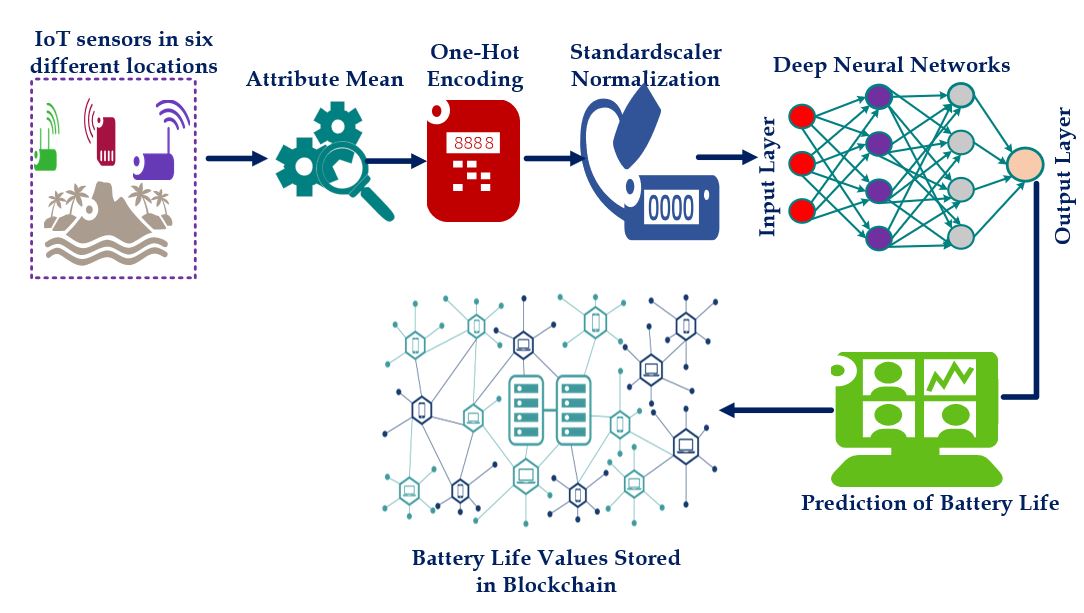}
	\caption{Proposed system architecture.}
	\label{fig:Framework-DNN-Blockchain} 
\end{figure}

The IoT network which is deployed in any real time environment will continuously sense some required data via the sensors. These sensor values are transmitted to its network storage for further decision making. The proposed architecture is as depicted in the Figure ~\ref{fig:Framework-DNN-Blockchain} and the steps of the proposed model is shown in Algorithm 1.



\begin{algorithm}[h]
	\SetAlgoLined
	Assign network id
	Initialize sensors network id\\
	 \For{sensor network id}
	 {
	 collect(sensor data sd)\\
	 \If{missing sd}
	 {
	 	run Attributemean(sd)
 	}
 	
	}
	Run One-HotEncoding(sd)\\
	Convert categorical data to numerical data \\
	Run Standardscalar(sd)\\
	Normalize sensor data into values between 0 to 1 \\
	Initialise Predicatedbatterylife bl\\
	bl = Run DNN(sd);\\
	\hspace{.3cm} Store network id, bl, dateofprediction \hspace{.1cm} $\tilde{}$  \hspace{.1cm} blockchain;\\
	End;
	
	\caption{Steps of the proposed
		model.}
\end{algorithm}

\begin{figure}[h!]
	\centering
	\includegraphics[width=7cm]{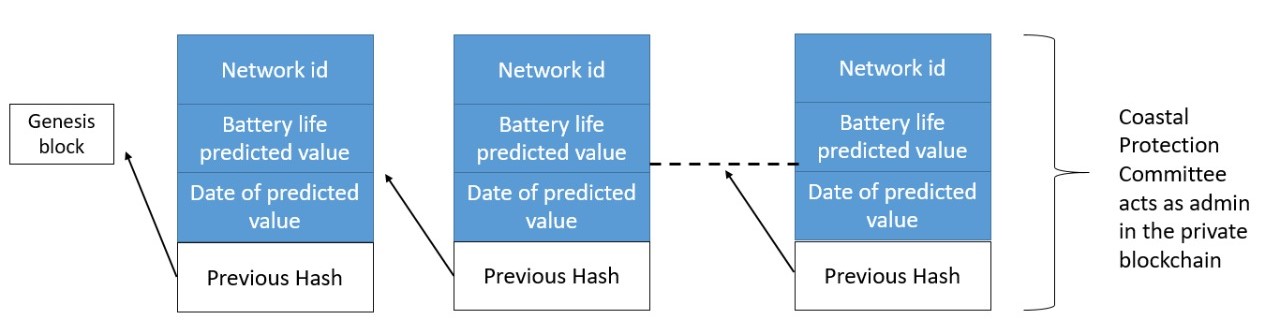}
	\caption{Private Blockchain Framework.}
	\label{fig:Blkarchnew} 
\end{figure}

Figure ~\ref{fig:Blkarchnew} shows the private blockchain framework where the coastal protection committee will have the authority for the data block creation and other stakeholders are brought into the blockchain through invites to view the data. Each data block contains three values namely: Network id – a unique number assigned to the set of sensors for identification, Battery life prediction – corresponds to the predicted battery life obtained by processing the sensor data through machine learning algorithms\cite{bhattacharya2020novel}, Date of predicted value – signifies the date on which the prediction has occurred. This value can be used to determine the time frame for battery exhaustion which in turn can be used to analyze the battery efficiency of the sensors.

The set of sensors in the IoT network will generate the data and this data will be processed by the machine learning algorithm to get the battery life prediction. This predicted value will then be uploaded to the blockchain. Any data uploaded to the blockchain cannot be tampered due to its immutability. 

 Hence, the  architecture as proposed in Figure ~\ref{fig:Blkarchnew} would be very feasible in securing the attained results.


\subsubsection{Distributed peer-peer network}
Blockchain is managed by a group of people in a distributed peer-peer network. When a new block is inserted into the chain, a copy of the block is echoed to all the nodes present in the network. Each node verifies the correctness of the block and adds it to its chain. The validity of the block is checked based on the consensus created by the nodes present in the network which tells about the guidelines, contracts and rules about classifying the block as a valid or invalid block. All these mentioned features make blockchain more secure than other technologies in the field.


\section{RESULTS AND DISCUSSIONS}
\label{sec5}

The results obtained for the real time beach water based IoT system in Chicago District are discussed. The methodology used for predicting the life of the battery in the IoT sensors and the security provided for storing the predicted battery life using blockchain technology is discussed in this section.




\begin{figure}[]
	\centering
	\includegraphics[width=7cm]{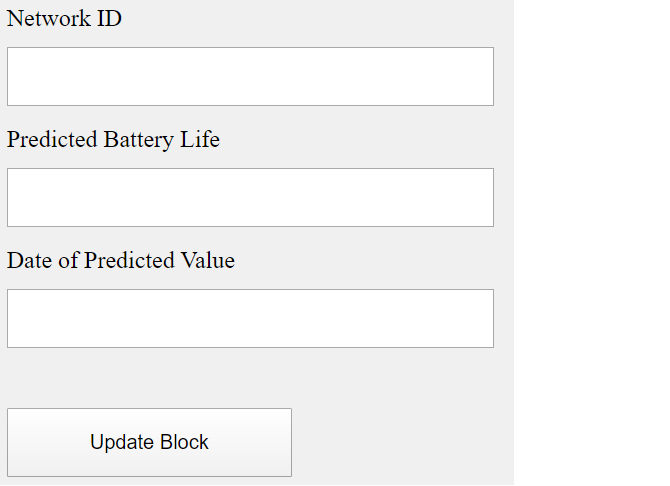}
	\caption{Admin side UI for block creation.}
	\label{fig:adminnew1} 
\end{figure}

\begin{figure}[]
	\centering
	\includegraphics[width=8cm]{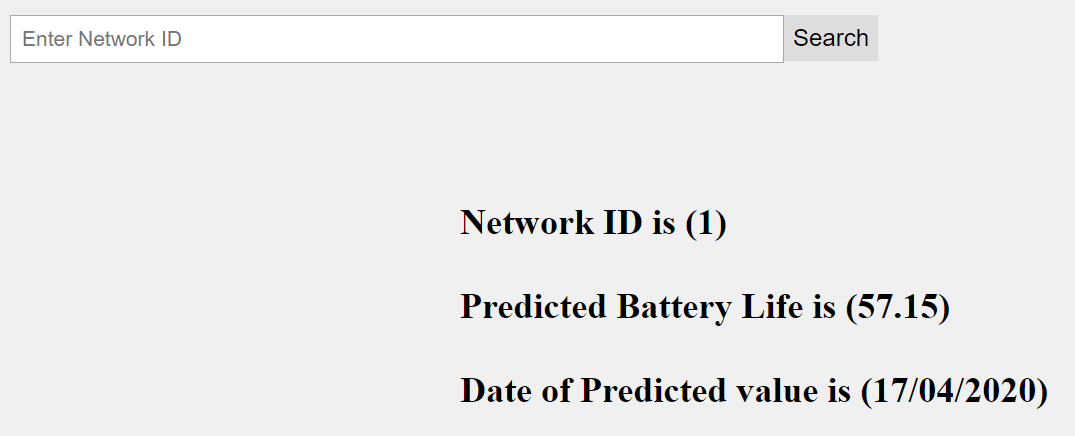}
	\caption{Stakeholders side UI.}
	\label{fig:user_new} 
\end{figure}

\subsection{Data Set Description}
The prediction model designed for predicting the battery life of IoT sensors that is deployed in the beach area of Chicago District is evaluated using the publicly available data set called "Chicago Park District Beach Water Quality Automated Sensors Data". There are six sensors that are deployed at the beach for monitoring the impact of waves. Since the sensors work on batteries, there is a chance that the battery drains out and the data is lost. Hence it is necessary to predict the life of battery well in advance and report to the Chicago Government based on various attributes and parameters. The data set which is used to support this prediction model has several features like name of th ebeach, water temperature, turbidity, transduder depth, period of wave, height of the wave, battery life (class label).

\subsection{Proposed Prediction Model and Results}

The prediction model that is proposed is based on Deep Neural Networks \cite{gadekallu2020early,24}. Deep Neural Networks is much better as compared to other Machine Learning models due to the availability of high performance computational systems such as Graphical Processing Units. There is one input layer, five layers that are hidden and finally one layer which forms the output. There are a total of 128 neurons in the first layer which forms the input layer. Since only one output is expected, that is the battery life, one neuron is used for this layer. The hidden layers vary from 64 to 256 neurons. The DNN is activated using the Relu Activation Function and the parameters required for the execution of the model are optimized using the ADAM Optimizer.

The proposed system secures the predicted battery life through the blockchain in a trust based manner. The admin updates the records with the help of UI depicted in Figure ~\ref{fig:adminnew1}. The predicted battery life along with the network ID and date are updated into the UI. Any stakeholder may be able to view this data block transacted to the blockchain. The user side search UI as shown in Figure  ~\ref{fig:user_new} may be used to obtain data of sensors in the IoT network by searching the block with its network ID. A set of all datablock related to that ID will be shown as a list.

\subsection{Performance Analysis}
\begin{figure}[]
	\centering
	\includegraphics[width=8cm]{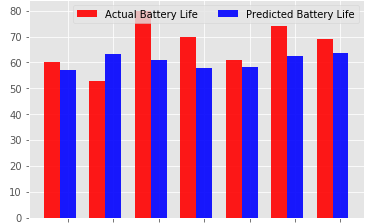}
	\caption{Predicted Battery Life vs Actual Battery Life.}
	\label{fig:DNN-Beach Water-25 instances} 
\end{figure}

The proposed prediction model based on DNN is compiled for 7 random instances and the result obtained is plotted in Figure ~\ref{fig:DNN-Beach Water-25 instances}. The battery life of the sensors which is predicted using the Proposed DNN based Prediction Model for seven random instances are tabulated in the Table ~\ref{Table2}. When compared to the actual battery life, the predicted battery life has an average accuracy of 90\%.

\begin{table}[h!]
	\centering
	\caption{Comparison of Actual Vs Predicted Battery Life using DNN }
	\label{Table2}
	\begin{tabular}{|p{80pt}|p{105pt}|}
		\hline
		Actual Battery Life & 
		Predicted Battery Life\\
		\hline
		60                           & 57.15    \\    	\hline                 
		53                           & 63.34                           \\ 	\hline
		80                           & 60.95                           \\ 	\hline
		70                           & 57.98                           \\ 	\hline
		61                           & 58.30                           \\ 	\hline
		74                           & 62.39                           \\ 	\hline
		69                           & 63.84                           \\ 	
		\hline
	\end{tabular}
\end{table}

\subsection{Insight of the Predicted Value}
The prediction model predicts the life of the battery of sensors deployed in the beaches of the Chicago District in six locations and the predicted value will be stored in the Blockchain which is publicly available and the government accesses the data whenever required for replacing the batteries at the right time as well as to place orders to the corporate companies who manufacture the batteries. 

The prediction accuracy is increased by 12\%. The proposed prediction model is evaluated using the parameters like Mean Absolute Error(MAE), Mean Squared Error(MSE), Root Mean Squared Error(RMSE) and Test Variance Score(TVS). The obtained results are tabulated in the Table 4. The tabulated results clearly depict that the MAE is 5.17, MSE is 35.70, RMSE is 6.02 and TVS is 0.19 in case of the proposed DNN Prediction Model which outperforms the other state of art techniques namely XGBoost and Linear Regression. Any prediction model is considered to be having high performance when the variance is low. From the results, it is evident that the variance score is lesser in DNN when compared to the other techniques.
\begin{table}[h!]
	\centering
	\caption{The Evaluation of Prediction Models using Standard Metrics}
	\label{Table3}
	\begin{tabular}{|p{40pt}|p{25pt}|p{45pt}|p{28pt}|p{40pt}|}
		\hline
		Model Used & Mean Absolute Error & Mean Squared Error & Root Mean Squared Error & Test Variance Score \\ \hline
		
		DNN (Proposed) & 5.17 & 35.70 & 6.02 & 0.19 \\ \hline
		
		Linear Regression & 988.72 & 1681700169.89 & 41008.53 & -25876324.98 \\ \hline
		
		XGBoost & 5.38 & 45.23 & 6.72 & 0.30 \\ \hline
	\end{tabular}
\end{table}

\subsection{Significance of using blockchain}
The major challenge of the existing centralized storage is that the data is owned by a single entity/stakeholder. This entity will have to provide access permission to view the data. Here, the data in question which is the battery life predicted values may be falsified to provide an unfair advantage. This issue arises due to a lack of trust which can be solved through the use of the distributed ledger of the blockchain\cite{25}. Here every stakeholder will obtain a copy of this data. This means falsification of this data by one entity is not possible since the change to be valid, needs to be updated in every ledger owned by other entities. This establishes trust\cite{26} over the current state of data. The data hash of the current block in the blockchain is referred by its succeeding block, which means any change to the block data can be identified through the change in its hash. This provides the basis for immutability in the blockchain which makes the system foolproof.  
\subsection{Real time use case for the proposed system}
Let us consider an example where remote IoT sensors are deployed deep into a forest for detection of forest fires. These senors are equipped with batteries left as stand alone units or as network groups. Once deployed, they process real time data and send back alerts regarding the same. If the  battery for this sensor falls short and requires an unpredicted maintenance, the time taken to replace them would result in loss of crucial data. The loss of this data could potentially mean a forest fire of an area goes undetected, the result of which could be catastrophic. This is such one such example where the result of unpredicted battery replacement could be hazardous. By combining this prediction of battery life with blockchain, this creates an immutable sharable copy of the battery life data of all the sensors to the necessary stakeholders with complete surety of the shared data.
The proposed system can be applied in similar scenarios where lack of such prediction mechanism may result in severe loss during business process. 
\section{Conclusion and Future Work}
\label{sec6}
The research is carried out and evaluated using the Chicago Park District Beach Water Data Set which is collected from a real-time IoT network deployed at six locations on the beaches.  The implementation focuses on predicting the battery life of the deployed sensors during various time frames by using the DNN based Prediction Model. The proposed  model outperforms the other available Machine Learning Models by an increase in prediction accuracy of 12\%. This system uses blockchain technology as its backend storage to store the predicted battery life values. This creates an immutable data record that can be viewed by all the stakeholders. If the data is made tamper-proof, the biggest concern regarding the trust of the delivered data can be mitigated. This combination of IoT coupled with blockchain can lead to newer possibilities in numerous fields such as supply chain management where real-time tracking of various parameters may be updated live to the stakeholders with a surety of untampered data. By reducing human intervention within the existing methodologies, a potent future for automation may be realized sooner using this combination.  
Though blockchain has several advantages, it has some unsolved research challenges such as scalability and delay in transaction processing which maybe addressed by future researchers.

\vspace{12pt}

\end{document}